\documentstyle[prl,aps,psfig]{revtex}

\begin{document}
\draft

\twocolumn[\hsize\textwidth\columnwidth\hsize\csname@twocolumnfalse\endcsname

\title{Electron correlation and the phase diagram of Si } 

\author{D. Alf\`{e}$^{1,2}$ and M. J. Gillan$^2$}

\address{
$^1$Geological Science Department, University College London 
Gower Street, London WC1E~6BT, UK \\
$^2$Physics and Astronomy Department, University College London 
Gower Street, London WC1E~6BT, UK}

\maketitle

\begin{abstract}
Previous first-principles calculations of the melting properties of
Si, based on the local-density approximation (LDA) for electronic
exchange-correlation energy, under-predict the melting temperature by
$\sim 20$~\%. We present new first-principles results
demonstrating that this problem is due to non-cancellation of
exchange-correlation errors between the semiconducting solid and the
metallic liquid. It is shown that other sources of error, particularly
those due to system size and Brillouin-zone sampling, can be made
negligible. The same LDA errors cause an underprediction of the
pressure of the diamond-Si~$\rightarrow$ beta-tin-Si transition.  The
generalized-gradient approximation largely corrects both features of
the Si phase diagram.
\end{abstract}

]

The long-standing ambition of calculating phase diagrams from
first-principles quantum mechanics has become a reality in the last 10
years~\cite{iron_melt,vocadlo00,vocadlo02,fppd_others}. An important
stimulus to the recent developments was the paper of Sugino and Car
(hereafter SG) on the melting of Si~\cite{sugino95}. The authors
showed how the technique of thermodynamic integration~\cite{frenkel96}
combined with first-principles molecular dynamics (FPMD)~\cite{car85}
based on density functional theory (DFT) can be used to calculated the
free energy of solids and liquids, and hence melting curves, with no
experimental input apart from fundamental constants. But although
their paper was influential, their numerical results on Si were not
very satisfactory, since their predicted melting temperature ($T_{\rm
m} = 1350$~K) was $\sim$~20~\% below the experimental value
(1685~K)~\cite{crc97}. Our purpose here is to identify the cause of
this discrepancy, which we shall argue comes from non-cancellation of
DFT errors between the solid and liquid phases, and specifically from
errors of the local-density approximation (LDA) used by SG. This has
implications for the reliability of other first-principles work on
phase diagrams.

The basic approximation in any DFT calculation is the algorithm
adopted for exchange- correlation energy $E_{\rm xc}$. Provided one
can eliminate all other sources of error in calculating total energies
and doing the statistical mechanics, then failure to reproduce
experimental melting properties must be due to errors in $E_{\rm
xc}$. But it is often claimed, even by first-principles practitioners,
that these other sources of error cannot be made small enough; in
particular, it is claimed that first-principles calculations cannot
yet be performed on large enough systems to render size errors
negligible~\cite{fppd_others}. This was one of the major issues
addressed by SG, who made strenuous efforts to ensure that their
non-$E_{\rm xc}$ errors were negligible; the results presented later
indicate that they were largely successful. Turning to $E_{\rm xc}$
errors, the crucial question is the extent to which they cancel
between the coexisting phases. Since diamond-structure Si (d-Si) is a
four-fold coordinated semiconductor and liquid Si (l-Si) is an
approximately six-fold coordinated metal~\cite{Stich89}, electron
screening is likely to be very different in the two phases, so that
non-cancellation of $E_{\rm xc}$ errors becomes an important issue. In
considering this, we are helped by the fact that the
pressure-stabilized $\beta$-tin structure ($\beta$-tin-Si) closely
resembles the liquid in being metallic and six-fold coordinated. This
suggests that there should be a close relation between the effect of
$E_{\rm xc}$ errors on the melting temperature and on the
d-Si~$\rightarrow$ $\beta$-tin-Si transition pressure, and an analysis
of this relation will help us to confirm that errors in the LDA
representation of $E_{\rm xc}$ account for the under-prediction of
$T_{\rm m}$.

Our first-principles calculations employ Vanderbilt ultra-soft
pseudopotentials~\cite{vanderbilt90} and plane-wave basis sets. Most
of our calculations are based on the local-density approximation (LDA)
for $E_{\rm xc}$ used by SG, but we shall also present results using
the generalized-gradient approximation (GGA)~\cite{wang91}. The
calculations were done with the VASP code~\cite{kresse96}. The
plane-wave cut-off was 150~eV, which gives a convergence of 6 meV/atom
in the difference of total (free) energies between liquid and solid,
and the pseudopotential core radii were 1.31 \AA. Our strategy for
computing the free energies of solid and liquid differs somewhat from
that of SG, and closely resembles that used in our recent work on
Fe~\cite{iron_melt} and Al~\cite{vocadlo02}.

The Helmholtz free energy $F$ of the solid can be written as $F =
F_{\rm perf} + F_{\rm vib}$, where $F_{\rm perf}$ is the free energy
of the perfect non-vibrating crystal (it is a {\em free} energy,
because we allow for thermal electronic excitations), and $F_{\rm
vib}$ is the contribution from lattice vibrations. The latter is
written as $F_{\rm vib} = F_{\rm harm} + F_{\rm anharm}$. The harmonic
free energy per atom $F_{\rm harm}$ in the classical limit (melting
occurs well above the Debye temperature) is: $F_{\rm harm} = 3 k_{\rm
B} T \ln ( \hbar \bar{\omega} / k_{\rm B} T )$, where the
geometric-mean frequency $\bar{\omega}$ is given by:
\begin{equation}
\ln ( \bar{\omega} ) = N_{{\bf k} s}^{-1} \sum_{{\bf k} s} \ln ( \omega_{{\bf k} s} ) \; ,
\end{equation}
with the sum going over wavevectors ${\bf k}$ and branches $s$ in the
Brillouin zone, $N_{{\bf k} s}$ being the number of terms in the
sum. The phonon frequencies $\omega_{{\bf k} s}$ are calculated using
the small-displacement method.~\cite{phonon}

The anharmonic contribution $F_{\rm anharm}$ turns out to be very
small (typically $\sim 15$~meV/atom near the melting temperature), so
that it is accurately given by the second-order expansion:
\begin{equation}
F_{\rm anharm} \simeq \langle U_{\rm anharm} \rangle_{\rm harm} -
\langle U_{\rm anharm}^2 \rangle_{\rm harm} / 2 k_{\rm B} T \; ,
\label{eqn:zwanzig}
\end{equation}
where $U_{\rm anharm}$ is the anharmonic part of the first-principles
total energy, and the thermal averages $\langle \, \cdot \,
\rangle_{\rm harm}$ are evaluated in the canonical ensemble of the
first-principles harmonic system. We have verified the accuracy of
Eq.~(\ref{eqn:zwanzig}) by comparing it with the exact expression:
$F_{\rm anharm} = - k_{\rm B} T \ln \langle \exp ( - U_{\rm anharm} /
k_{\rm B} T ) \rangle_{\rm harm}$.

Our calculations of $F_{\rm perf}$ were performed on the primitive
two-atom unit cell, at volumes from 16 to 22~\AA$^3$/atom with
$k$-point sampling dense enough to give a precision of $\sim
0.1$~meV/atom. Results were fitted to the
Birch-Murnaghan~\cite{birch47} form, which reproduces the data to
within $\sim 0.1$~meV/atom. For $F_{\rm harm}$, we calculated the
force-constant matrix using 54-atom cells, with spot-checks on cells
of up to 250 atoms indicating convergence to within $\sim
2$~meV/atom. The calculations were done at volumes from 18 to
21~\AA$^3$/atom, and $\ln ( \bar{\omega} )$ was fitted to a
second-order polynomial $\ln ( \bar{\omega} ) = a + b V + c V^2$,
which gives a fitting error in $F_{\rm harm}$ of $\sim
1$~meV/atom. The thermal averages needed to calculate $F_{\rm anharm}$
were done on a 54-atom cell at volumes of $V = 18$ and 20~\AA$^3$/atom
and temperatures of 1000, 1500 and 2000~K. The results are accurately
reproduced by the quadratic form $F_{\rm anharm} = a T^2$, and to the
accuracy we require it is enough to take the value $a = 7 \times
10^{-9}$~eV~K$^{-2}$ for both volumes.

The free energy of the liquid is calculated using thermodynamic
integration (TI), with the Stillinger-Weber~\cite{stillinger85}
empirical total-energy model used as reference system. The difference
of Helmholtz free energy $\Delta F \equiv F_{\rm AI} - F_{\rm ref}$
between the {\em ab initio} and Stillinger-Weber systems is obtained
using the standard formula:
\begin{equation}
\Delta F = \int_0^1 d \lambda \, \langle U_{\rm AI} - U_{\rm ref} \rangle_\lambda \; ,
\label{eqn:ti}
\end{equation}
with $U_{\rm AI}$ and $U_{\rm ref}$ the {\em ab initio} and reference
total-energy functions, and $\langle \, \cdot \, \rangle_\lambda$ the
thermal average evaluated in the ensemble of the system whose
total-energy function is $U_\lambda \equiv ( 1 - \lambda ) U_{\rm ref}
+ \lambda U_{\rm AI}$. In practice, the integral over $\lambda$ is
performed either by evaluating $\langle U - U_{\rm ref}
\rangle_\lambda$ at a set of $\lambda$ values and using Simpson's
rule, or by using `adiabatic switching', in which $\lambda$ is slowly
and continuously varied between the two limits~\cite{watanabe90}. The
reference free energy $F_{\rm ref}$ is calculated by thermodynamic
integration starting from the Lennard-Jones system, for which accurate
free energies have been published~\cite{johnson93}. The calculation of
$F_{\rm ref}$ is done on very large systems, so that it is converged
with respect to system size to better than 1~meV/atom.

As has often been stressed~\cite{iron_melt}, the final results for
$F_{\rm AI}$ do not depend on the choice of reference system, but the
{\em efficiency} of the calculations can be greately improved by
careful tuning of the reference system, the criterion being that the
strength of the fluctuations of $U_{\rm AI} - U_{\rm ref}$ should be
made as small as possible. We find that in this sense the original
parameters of the Stillinger-Weber model are far from optimal for
liquid Si. Using {\em ab initio} MD simulations of l-Si at the state
$V = 18.16$~\AA$^3$/atom, $T = 2000$~K, we have varied the
Stillinger-Weber parameters to minimise the fluctuation strength, and
it is the resulting `optimized' SW model that we use as our reference
model.

We made thorough tests of the convergence of $\Delta F$ with respect
to system size and electronic $k$-point sampling by calculating it at
the representative state point $V = 17$~\AA$^3$/atom and $T = 1750$~K,
using systems of up to 512 atoms and up to 36
Monkhorst-Pack~\cite{monkhorst76} $k$-points (results of these tests
in Table~1). The tests were done as follows. The $\Gamma$-point
results were obtained by explicit simulations on systems of all sizes,
with $\Delta F$ calculated by thermodynamic integration
(Eq.~(\ref{eqn:ti})). In most cases, we used the five $\lambda$ values
0.0, 0.25, 0.5, 0.75 and 1.0 together with Simpson's rule, and
comparisons with other sets of $\lambda$ values show that the residual
error from the integration itself is less than 5~meV/atom. We then
used thermodynamic integration, with the $\Gamma$-point system as the
reference system, to obtain the results for other $k$-point
samplings. For systems of $N \ge 128$ atoms, the fluctuations of the
difference of energies calculated with $\Gamma$-point and more
$k$-points is small enough to allow the second-order expansion
to be used instead of explicit TI, but for $N = 64$ this is not
adequate and we used explicit TI.

The results of Table~1 show that with 64 atoms and four $k$-points the
free-energy difference $\Delta F$ between {\em ab initio} and
optimized Stillinger-Weber is converged to better than 10~meV/atom,
and we have used this system to obtain $F_{\rm AI}$ for the liquid at
the set of state points $V = 16$, 17, 18, 19 and 20~\AA$^3$/atom and
$T = 1250$, 1500 and 1750~K. At each $T$, $F_{\rm AI}$ was fitted to a
Birch-Murnaghan equation of state, the residual fitting error being no
more than 2~meV/atom.

Our fitted {\em ab initio} Helmholtz free energies of d-Si and l-Si
allow us to obtain the Gibbs free energy $G \equiv F - V ( \partial F
/ \partial V )_T$, and hence the melting curve.  The zero-pressure
results for $T_{\rm m}$ and the entropy and volume of fusion, $\Delta
S$ and $\Delta V$, are compared in Table~2 with those of SG and the
experimental values. Our very close agreement with the SG value of
$T_{\rm m}$ (difference of only 50~K) confirms that their size and
$k$-point errors were indeed very small, and also confirms that LDA
under-predicts $T_{\rm m}$ by $\sim 20$~\%. We note that our $\Delta
S$ and $\Delta V$ values are both somewhat greater than those of SG.

We now turn to the matter of non-cancelling LDA errors between phases,
exploiting the electronic and structural similarity between d-Si and
$\beta$-tin-Si. At room temperature, the transition d-Si~$\rightarrow$
$\beta$-tin-Si occurs at an experimental pressure in the range $10.3 -
12.5$~GPa~\cite{experiments} (although also a low value of 8.8 GPa has
been reported~\cite{olijnyk84}).  Earlier LDA calculations on the
static zero-temperature crystals gave transition pressures in the
range $7.8 - 8.4$~GPa~\cite{boyer91,moll95}, and our own calculations
yield the value 7.8~GPa, which agrees closely with the earlier
values. However, it has been shown that temperature has a strong
influence on the transition pressure, which drops by $\sim 20$~\% as
$T$ goes from 0~K to room temperature~\cite{gaal-nagy01}, so that the
temperature-corrected LDA pressure is too low by at least 4~GPa. It is
also known that the generalized-gradient approximation (GGA) for
$E_{\rm xc}$ significantly improves the predicted transition pressure.
With the Perdew-Wang GGA~\cite{wang91}, we find a transition pressure
of 11.7~GPa (9.4~GPa when corrected to room temperature), which agrees
closely with earlier GGA values~\cite{moll95}. Our calculations show
that the main reason why LDA under-predicts the transition pressure is
that it erroneously shifts the energy of d-Si upwards relative to
$\beta$-tin-Si. The GGA goes a long way towards correcting this
destabilization of d-Si.  But a low melting temperature is also a sign
of an erroneous destabilization of d-Si, and we hypothesize that the
same underlying $E_{\rm xc}$ error is responsible for both
under-predictions.

To test this hypothesis, we have recalculated the melting properties
using GGA. It is instructive to do this by evaluating the free energy
difference between the LDA and GGA systems. We have therefore
performed long simulations for solid and liquid at the zero pressure
volumes using the LDA, and calculated the GGA energies at a number of
statistically independent configurations, for both the solid and the
liquid. The calculations have been done on cells containing 64 atoms
with four $k$-points, and spot-checked with calculations on cells
containing 512 atoms and $\Gamma$-point sampling.  Firstly, we found
that the energy differences between GGA and LDA are basically
constant, i.e. do not depend on the configurations of the atoms, which
confirms the idea that the shift should be the same as for the low
temperature static lattices. Secondly, we found that the free energy
of the liquid is raised by 88~meV/atom relative to that of d-Si.
Given an LDA entropy change on melting of $3.5 k_{\rm B}$/atom, it is
easy to work out a shift of melting temperature GGA-LDA of 292~K,
bringing the GGA result to 1590~K, in much closer agreement with the
experimental datum. We also found that, at the volumes corresponding
to the LDA zero pressure, the GGA pressures are about 3.5 GPa larger
than the LDA ones, so the GGA zero pressure volumes are
larger. However, the bulk moduli for the solid and the liquid at the
melting temperature are 78 and 34 GPa respectively, so the liquid will
expand more than the solid in the GGA.  We can estimate a new volume
change on melting of 9.4\%, which is also in somewhat better agreement
with the experiments.

In summary, we have shown that the key issue in a first-principles
account of the melting properties of Si is non-cancellation of
exchange-correlation errors between solid and liquid because of their
different electronic structure. Technical errors due to system size
and $k$- point sampling are readily brought under tight control. The
basic reason why this can be done is that system size affects only the
small difference of free energy between the first-principles system
and a carefully designed reference system. The non-cancellation of
exchange-correlation errors between coexisting semiconductor and metal
is also responsible for difficulties in predicting the pressure of the
diamond-Si~$\rightarrow$ $\beta$-tin-Si transition, and there is a
quantitative relation between the error in this transition pressure
and the error in melting temperature. The general implication is that
for phase equilibria in which the coexisting phases have essentially
the same electronic structure, e.g. the melting of high- pressure Fe,
DFT calculations can be expected to predict phase equilibria with
satisfactory accuracy.

\section*{Acknowledgments}
The work of DA is supported by a Royal Society University Research
Fellowship. The calculations were performed at the UCL HiPerSPACE
Centre, supported by HEFCE grant JR98UCGI and EPSRC grant
GR/R38156. We acknowledge valuable discussions with R.~Needs.

\newpage
\begin{table}
\begin{tabular}{lccccc}
\hline
$N$ & $\Delta F_1$ & $\Delta F_4$ & $\Delta F_8$  & $\Delta F_{32}$ & $\Delta F_{36}$ \\
\hline
64   &  -4.165(5)  & -4.262(5)  &  -4.253(5) &  -4.257(5) & -4.257(5) \\
128  &  -4.282(5)  & -4.250(5)  &            &            &           \\       
216  &  -4.281(5)  & -4.262(5)  &            &            &           \\     
512  &  -4.248(5)  & -4.251(5)  &            &            &     
\end{tabular}
\caption{Difference $\Delta F$ of Helmholtz free energy (in units of
eV/atom) at thermodynamic state $V=17$ \AA$^3$/atom, $T=1750$ K,
between the {\em ab-initio} and the Stillinger-Weber potential as
function of size of simulated system (number of atoms $N$) and number
of Monkhorst-Pack ${\bf k}$-points (subscript on $\Delta
F$).}\label{tab:size}
\end{table}

\begin{table}
\begin{tabular}{lcccc}
\hline
& This work (LDA) & This work (GGA) & Sugino and Car~\protect\cite{sugino95} & Experiment \\
\hline
$T_m$(K) & 1300(50) & 1590(50) & 1350(100) & 1685(2)$^a$ \\
$\Delta V_m/V_s$ & 0.142 & 0.094 & 0.1 & 0.119$^b$, 0.095$^c$ \\
$\Delta S_m$ & 3.5 &  & 3.0 & 3.6$^d$,3.3$^c$\\
$dT_m/dP$ & -58 &  & -50  &  -38$^a$\\
\hline
\end{tabular}
$^a$ Ref.~\protect\cite{crc97}\\
$^b$ Ref.~\protect\cite{gabathur79} \\
$^c$ Ref.~\protect\cite{ubbelohde78} \\
$^d$ Ref.~\protect\cite{barin73}
\caption{Comparison of calculated and experimental melting properties
of Si at ambient pressure: melting temperature $T_m$, volume change
$\Delta V_m$ divided by volume of solid at melting temperature,
entropy change $\Delta S_m$ per atom divided by Boltzmann's constant,
and slope of melting curve $dT_m/dP$ (units of
K~GPa$^{-1}$).}\label{tab:results}
\end{table}

\end{document}